\PassOptionsToPackage{colorlinks=true,citecolor=blue,filecolor=blue,linkcolor=red,urlcolor=red}{hyperref}
\documentclass[sigconf,9pt]{acmart}

\setcopyright{acmlicensed}

\acmSubmissionID{146}
\copyrightyear{2022}
\acmYear{2022}
\setcopyright{acmcopyright}
\acmConference[ICS '22]{2022 International Conference on Supercomputing}{June 28--30, 2022}{Virtual Event, USA}
\acmBooktitle{2022 International Conference on Supercomputing (ICS '22), June 28--30, 2022, Virtual Event, USA}
\acmPrice{15.00}
\acmDOI{10.1145/3524059.3532376}
\acmISBN{978-1-4503-9281-5/22/06}

\usepackage[utf8]{inputenc}
\usepackage[T1]{fontenc}
\usepackage[english]{babel}
\usepackage{amsmath,amsfonts,amsthm}
\usepackage{textcomp}
\usepackage{booktabs}
\usepackage[autostyle]{csquotes}
\usepackage[printonlyused]{acronym}
\usepackage{multirow}
\usepackage{xspace}
\usepackage{subcaption}
\usepackage[colorlinks=true,citecolor=blue,filecolor=blue,linkcolor=red,urlcolor=red]{hyperref}
\usepackage{tikz}
    \usetikzlibrary{calc}
\usepackage{pgfplots}
    \pgfplotsset{compat=1.14}

\addto\extrasenglish{%
}


\def\anyseq{AnySeq/GPU\xspace}%

\newacro{AST}{abstract syntax tree}
\newacro{DP}{dynamic programming}
\newacro{GCUPS}{giga cell updates per second}
\newacro{NGS}{next-generation sequencing}
\newacro{HLS}{high-level synthesis}
\newacro{SPMD}{single program, multiple data}
\newacro{TCUPS}{tera cell updates per second}

\def\impala{Impala\xspace}

\usepackage{listings}
\usepackage{lstautogobble}
\usepackage{etoolbox}

\expandafter\patchcmd\csname \string\lstinline\endcsname{%
    \leavevmode
    \bgroup
}{%
    \leavevmode
    \ifmmode\hbox\fi
    \bgroup
}{}{%
    \typeout{Patching of \string\lstinline\space failed!}%
}

    {\noindent\ignorespaces\begin{lstlisting}[#1,float,floatplacement=H]}{\end{lstlisting}\noindent\ignorespacesafterend}

\definecolor{whitesmoke}{rgb}{0.96, 0.96, 0.96}

\definecolor[named]{codegreen}    {named}{ACMGreen}
\definecolor[named]{codered}      {named}{ACMRed}
\definecolor[named]{codelightblue}{named}{ACMLightBlue}
\definecolor[named]{codedarkblue} {named}{ACMDarkBlue}

\def\lst{\lstinline}

\makeatletter
\newenvironment{btHighlight}[1][]
{\begingroup\tikzset{bt@Highlight@par/.style={#1}}\begin{lrbox}{\@tempboxa}}
{\end{lrbox}\bt@HL@box[bt@Highlight@par]{\@tempboxa}\endgroup}

\newcommand\btHL[1][]{%
    \begin{btHighlight}[#1]\bgroup\aftergroup\bt@HL@endenv%
}
\def\bt@HL@endenv{%
    \end{btHighlight}%
    \egroup
}
\newcommand{\bt@HL@box}[2][]{%
    \tikz[#1]{%
        \pgfpathrectangle{\pgfpoint{1pt}{0pt}}{\pgfpoint{\wd #2}{\ht #2}}%
        \pgfusepath{use as bounding box}%
        \node[anchor=base west, fill=codelightblue,outer sep=0pt,inner xsep=1pt, inner ysep=0pt, rounded corners=3pt, minimum height=\ht\strutbox+1pt,#1]{\raisebox{1pt}{\strut}\strut\usebox{#2}};
    }%
}
\makeatother

\lstdefinestyle{node}{
    backgroundcolor=,
    language=,
    basicstyle=\tiny\ttfamily,
    morekeywords = {br,neg,or,and,all,any},
    numbers=none,
    mathescape=true,
    frame=none,
    literate={<-}{{$\leftarrow$}}1
}

\lstdefinelanguage{alg}{
    morecomment = [s]{/*}{*/},
    morecomment = [l]{//},
    sensitive = true,
    morekeywords = {for,next,to,step}
}

\lstdefinelanguage{impala}{
    morecomment = [s]{/*}{*/},
    morecomment = [l]{//},
    sensitive = true,
    morekeywords = {i8,i16,i32,i64,u8,u16,u32,u64,f16,f32,f64,bool,int,float,double,extern,struct,as,match,true,false,type,with,let,mut,while,in,exit,return,break,continue,if,else,for,do,fn,enum},
    moredelim=**[is][\btHL]{§}{§},
    morestring=[b]",
}

\lstdefinelanguage{metaocaml}{
    sensitive = true,
    morekeywords = {let,in,rec,if,then,else,fun},
    moredelim=**[is][\btHL]{§}{§},
}

\lstdefinelanguage{pseudoml}{
    sensitive = true,
    morekeywords={fun,where,whererec,lambda,let,letrec,in,and,bool,float,int,br,noret},
    literate=%
        {==}{{=}}1
        {!=}{{$\neq$}}1
        {<=}{{$\leq$}}1
        {>=}{{$\geq$}}1
        {->}{{$\rightarrow$}}1
        {<-}{{$\leftarrow$}}1
        {bot}{{$\bot$}}1
        {LAMBDA}{{$\lambda$}}1
}

\lstdefinelanguage{scala}{
    morecomment = [s]{/*}{*/},
    morecomment = [l]{//},
    sensitive = true,
    morekeywords = {val,var,new,with,import,trait,this,def,if,else,Int},
    moredelim=**[is][\btHL]{§}{§},
}

\lstdefinelanguage{scheme}{
    sensitive = true,
    morekeywords={define,filter}
}

\lstdefinelanguage{sierra}{
    morecomment = [s]{/*}{*/},
    morecomment = [l]{//},
    morestring=[b]",
    sensitive = true,
    morekeywords = {uniform,varying,simd,scalar,for_each_active,for_each_unique,current_mask},
    morekeywords = {kernel,uint,mask,skip,true,false,uint32_t,uint64_t,nullptr,return,public,protected,private,template,auto,class,virtual,struct,union,void,this,size_t,volatile,if,else,do,while,case,goto,switch,for,while,bool,typedef,static,const,float,int,short,char,double,break,continue},
    keywords = {[2]define},
    keywordstyle={[2]\color{uds-purple}\bfseries},
    moredelim=**[is][\btHL]{§}{§},
}

\lstdefinelanguage{ssa}{
    sensitive = true,
    morekeywords={fn,bool,float,int,phi,goto,br,return},
    literate=
        {:=}{{$\gets$}}1
        {==}{{=}}1
        {!=}{{$\neq$}}1
        {<=}{{$\leq$}}1
        {>=}{{$\geq$}}1
        {->}{{$\rightarrow$}}1
        {<-}{{$\leftarrow$}}1
        {PHI}{{$\phi$}}1
}

\lstdefinelanguage{terra}{
    morecomment = [s]{/*}{*/},
    morecomment = [l]{//},
    sensitive = true,
    morekeywords = {int,function,if,then,return,else,elseif,terra,end,local},
    moredelim=**[is][\btHL]{§}{§},
}

\begin{document}

\title[AnySeq/GPU]{AnySeq/GPU: A Novel Approach for Faster Sequence Alignment on GPUs}

\author{Andr\'{e} M\"uller}
\email{{muelan,bertil.schmidt}@uni-mainz.de}
\author{Bertil Schmidt}
\affiliation{%
    \institution{Johannes Gutenberg University}
    \city{Mainz}\country{Germany}
}

\author{Richard Membarth}
\email{richard.membarth@thi.de}
\affiliation{%
    \institution{Technische Hochschule Ingolstadt (THI)\\Research Institute AImotion Bavaria}
    \city{Ingolstadt}\country{Germany}
}
\additionalaffiliation{%
    \institution{German Research Center for Artificial Intelligence (DFKI), Saarland Informatics Campus}
    \city{Saarbrücken}\country{Germany}
}
\author{Roland Leißa}
\email{leissa@uni-mannheim.de}
\affiliation{%
    \institution{University of Mannheim}
    \city{Mannheim}\country{Germany}
}
\author{Sebastian Hack}
\email{hack@cs.uni-saarland.de}
\affiliation{%
    \institution{Saarland University, Saarland Informatics Campus}
    \city{Saarbrücken}\country{Germany}
}

\begin{abstract}
%Many bioinformatics applications spend a significant portion of execution time on pairwise alignments of sequencing reads.
In recent years, the rapidly increasing number of reads produced by \ac{NGS} technologies has driven the demand for efficient implementations of sequence alignments in bioinformatics.
%most notably those based on dynamic programming like the Smith-Waterman algorithm.
%While many of the available alignment libraries are optimized for multi-core CPU systems, there exist still few implementations that can fully leverage the massively parallel processing capabilities of GPUs.
However, current state-of-the-art approaches are not able to leverage the massively parallel processing capabilities of modern GPUs with close-to-peak performance.

We present \anyseq---a sequence alignment library that
%supports batched pairwise alignments of short and long sequencing reads.
augments the AnySeq\,1 library with a novel approach for accelerating \ac{DP} alignment on GPUs by minimizing memory accesses using warp shuffles and half-precision arithmetic. Our implementation is based on the AnyDSL compiler framework which allows for convenient zero-cost abstractions through guaranteed partial evaluation.
We show that our approach achieves over 80\% of the peak performance on both NVIDIA and AMD GPUs thereby outperforming the GPU-based alignment libraries AnySeq\,1, GASAL\,2, ADEPT, and NVBIO by
a factor of at least 3.6 while achieving a median speedup of 19.2$\times$ over these tools across different alignment scenarios and sequence lengths when running on the same hardware.

%as well as CPU alignment libraries like SeqAn executed on many-core dual-CPU workstations.
This leads to throughputs of up to 1.7 TCUPS (tera cell updates per second) on an NVIDIA GV100, up to 3.3 TCUPS with half-precision arithmetic on a single NVIDIA A100, and up to 3.8 TCUPS on an AMD MI100.
\anyseq is publicly available at
\url{https://github.com/AnyDSL/anyseq}.

%\todo{If you have numbers like X\% of peak performance, I think they would be easier to understand a probably even more impressive than saying Y TCUPS.}
\end{abstract}

\maketitle
\acresetall

%==============================================================================
\section{Introduction}
%==============================================================================
 Computing alignments of genomic sequences is a core algorithmic component in bioinformatics. With the development of \ac{NGS} technologies, increasing amounts of high throughput sequencing read datasets are produced, which establishes the need for highly optimized alignment implementations.

% Recent years have seen a tremendous increase in the volume of data generated in the life sciences, especially propelled by the rapid progress of \ac{NGS} technologies.
% As a consequence, modern bioinformatics tools often require highly efficient implementations of core sequence analysis algorithms.

Pairwise alignment of two given genomic sequences aims to identify an optimal order-preserving mapping of their characters while allowing  insertions of gaps. Needleman-Wunsch~\cite{nw}, Smith-Waterman~\cite{sw} and their variants are frequently used algorithms to compute optimal local, global, or semi-global pairwise alignments, respectively, by means of \ac{DP}. However, their time complexity is proportional to the product of sequence lengths, thus making them highly time consuming for typical \ac{NGS} datasets which either consist of many short reads (of a few hundred base-pairs in length for Illumina platforms) or long reads (varying between a few thousands to hundreds of thousand base-pairs in length for the PacBio and ONT platforms).

% Given a pair of genomic sequences, a common operation in bioinformatics is to identify their similarity under a model of evolution which allows for certain sequence modifications.
%This leads to so-called \emph{sequence alignments} that map characters across the sequences in an order-preserving way while potentially inserting gaps such that a mathematical model of their similarity is maximized.
%For pairwise alignment computation, the Smith-Waterman algorithm~\cite{sw}, the Needleman-Wunsch algorithm~\cite{nw}, and their variants are widely used.
%These compute an optimal local, global, or semi-global alignment of two sequences under a given scoring scheme by means of \ac{DP}.
%However, the associated time complexity proportional to the product of sequence lengths makes this approach a time consuming component of various bioinformatics workflows.
%As a consequence, these algorithms have been optimized on numerous architectures including CPUs~\cite{swaphils, hou2016aalign, misra2018performance}, GPUs~\cite{cudalign, liu2013cudasw++, korpar2013sw, de2016cudalign}, and FPGAs~\cite{oliver2005hyper, fpga, rucci2018swifold}.

Consequently, a variety of parallelized implementations and libraries have
been developed for computing pairwise sequence alignments in recent years on CPUs~\cite{rahn2018generic,misra2018performance,parasail,ssw}, GPUs~\cite{cudalign, liu2013cudasw++, korpar2013sw, de2016cudalign, HouWFVL18, gasal2, ADEPT, nvbio}, and FPGAs~\cite{oliver2005hyper, fpga, rucci2018swifold}.
However, existing GPU implementations are limited by inefficient memory access schemes and thus cannot fully exploit the performance of modern GPUs. Furthermore, they are not performance portable to highly varying sequence lengths and different alignment types needed in \ac{NGS} bioinformatics workflows. In addition, they are only optimized for certain types of (CUDA-enabled) GPUs.
Thus, there is an urgent need for a flexible alignment library that reaches close-to-peak performance on modern GPUs from various vendors.

%------------------------------------------------------------------------------
%\subsection{Contributions}
%------------------------------------------------------------------------------

We address this need by presenting \anyseq---a highly efficient GPU-based extension of the AnySeq\,1~\cite{DBLP:conf/ipps/MullerS0MLKH20} library for batched pairwise alignments of short and long sequencing reads. Our implementation uses the AnyDSL compiler framework~\cite{DBLP:journals/pacmpl/LeissaBHPMSMS18}  which is based upon the concept of \emph{partial evaluation}~\cite{Futamura:1999:PEC:609149.609205,Consel,Brady} and allows for compilation of different variants of the \ac{DP} algorithm that are highly optimized for specific alignment types, scoring schemes, and hardware targets.

\subsection{Contributions}

This paper makes the following contributions:
\begin{itemize}
\item We present the design of a novel fine-grained parallelization strategy based on warp intrinsics for sequence alignment targeting massively parallel GPU architectures.
    We introduce a \ac{DP} matrix partitioning scheme that supports batches with highly varying sequence lengths (see \autoref{sec:mapping}).
\item We showcase how \anyseq uses state-of-the-art DSL technology which makes it possible to implement large parts of \anyseq in a hardware-independent way and instantiate these parts with highly-optimized kernels for AMD and NVIDIA GPUs (see \autoref{sec:kernelvariants}).
\item We demonstrate that our implementation can achieve over 80\% of the available peak performance on  modern GPUs from both NVIDIA and AMD for single- and half-precision arithmetic.
    For various types of alignments (local, global, semi-global, w/ and w/o traceback), gap penalties (linear, affine), and different \ac{NGS} technologies (Illumina, PacBio) \anyseq achieves speedups of at least 3.6$\times$ and a median speedup of 19.2$\times$ over state-of-the-art GPU codes (GASAL2, ADEPT, NVBIO, AnySeq\,1) executed on the same hardware (see \autoref{performance}).
\end{itemize}

% Plz don't have this mini table of contents. Readers will find the Conclusion section themselves Oo
%The rest of the paper is organized as follows. \autoref{overview} provides background on alignment, GPU computing, AnyDSL, and related work. Our parallelization scheme and its implementation are presented in \autoref{methods}. Performance is evaluated in \autoref{performance}. \autoref{conclusion} concludes the paper.

%------------------------------------------------------------------------------
\section{Background}\label{overview}
%------------------------------------------------------------------------------

%------------------------------------------------------------------------------
\subsection{Sequence Alignment}\label{alignment}
%------------------------------------------------------------------------------

%The optimum alignment of two sequences can be computed using dynamic programming (\ac{DP}).
%The conceptual layout of the \ac{DP} matrix and its data dependency on three neighboring cells are illustrated in \autoref{fig:dpmatrix}.
%Their optimal alignment can be found in $\mathcal{O}(m \cdot n)$ by recursively solving three smaller subproblems.
%optimal alignment score $H(i,j)$ for the prefixes $(q_1 \ldots q_i)$ and $(s_1 \ldots s_j)$ is given by the recurrence relation

\begin{figure}[t]
    \centerline{\includegraphics[width=0.35\textwidth]{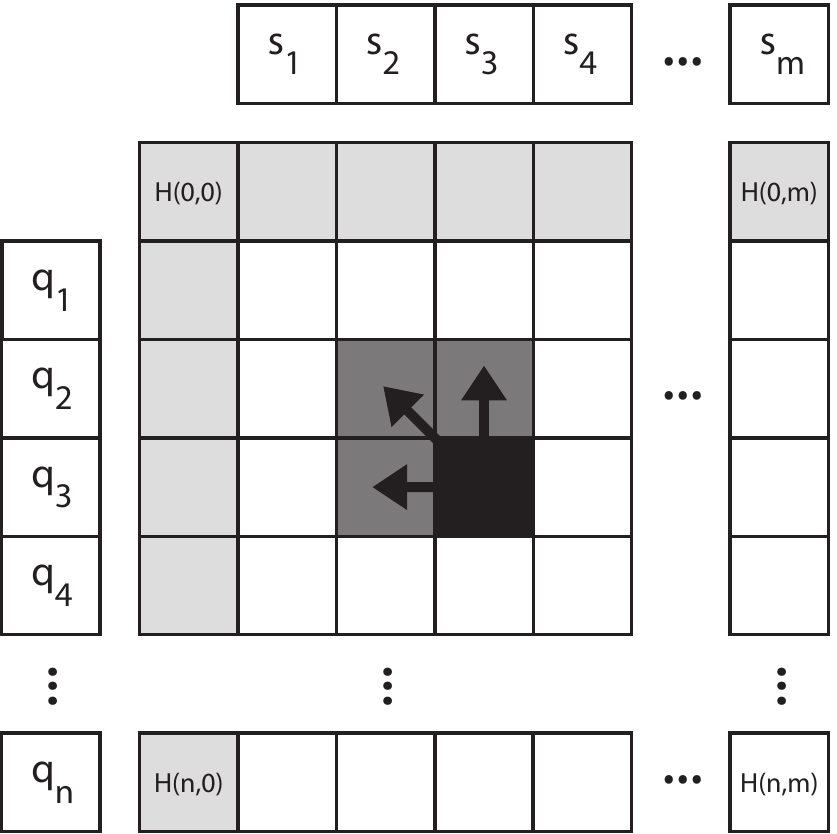}}
    \caption{The \ac{DP} matrix $H$}
    \label{fig:dpmatrix}
\end{figure}

Let $Q=(q_1 q_2 \ldots q_m)$ and $S=(s_1 s_2 \ldots s_n)$ be two genomic sequences over the alphabet $\Sigma = \{ A, C, G, T\}$.
For each pair $(q_i,s_j)$ of characters the \ac{DP} algorithm decides if these characters should be aligned or if a gap should be inserted and records a corresponding score in a matrix $H$. This is illustrated in \autoref{fig:dpmatrix}, where light gray cells are initialization cells and
dark gray cells indicate the ancestral subproblems of the currently active cell in black.
%The right image shows the cells that need to be stored for score-only computation in dark gray.
Formally, we define $H$ as a recurrence relation:
\begin{equation} \label{eq:H}
    H(i,j) = \max
    \begin{cases}
        H(i-1,j-1) + \sigma(q_i,s_j) \\
        E(i,j) \\
        F(i,j) \\
        \nu \\
    \end{cases}
    \quad
    \begin{array}{c}
        1 \leq i \leq m \\
        1 \leq j \leq n
    \end{array}
\end{equation}
where
\begin{itemize}
    \item the first row and column of $H$ are given as initialization,
    \item $E$ and $F$ determine the \emph{gap penalty scheme} (linear or affine),
    \item $\nu$ and the initialization model determine the \emph{alignment type} (local, global, or semi-global).
\end{itemize}
This yields the optimal alignment score $H(i,j)$ of the prefixes $(q_1 \ldots q_i)$ and $(s_1 \ldots s_j)$ where $\sigma$ is a substitution function over $\Sigma \times \Sigma$ that determines the score of aligning two characters.

\paragraph{Linear Gap Penalty}
For this gap penalty scheme, we just use a gap penalty~$\alpha$:
\begin{align}
    E(i,j) &= H(i-1,j) - \alpha \\
    F(i,j) &= H(i,j-1) -  \alpha
\end{align}
%\begin{align} \label{eq:lingap}
%    E(i,j) &= H(i-1,j) - g \\
%      F(i,j) &= H(i,j-1) - g
%\end{align}

\paragraph{Affine Gap Penalty}
For this variant, we penalize a gap of length~$k$ by $\alpha + (k-1) \cdot \beta$ where $\alpha$ is the cost of the first gap (opening) and $\beta$ the cost of each subsequent gap (extension):
%The following functions determine the best alignment score for the prefixes $(q_1 \ldots q_i)$  and $(s_1 \ldots s_j)$ under the constraint that $s_j$ (or $q_i$) is aligned to a gap:
%and the score of the best alignment for $(q_1 \ldots q_i)$ and $(s_1 \ldots s_j)$ under the constraint that $q_i$ is aligned to a gap becomes
\begin{align}
\label{eq:E}
    E(i,j) &= \max
    \begin{cases}
        E(i-1,j) - \beta \\
        H(i-1,j) - \alpha
    \end{cases} \\
\label{eq:F}
    F(i,j) &= \max
    \begin{cases}
        F(i,j-1) - \beta \\
        H(i,j-1) - \alpha.
    \end{cases}
\end{align}

%Initialization of the first rows and columns of $H$, $E$, and $F$---as well as in what cell(s) to look for the optimal score---depends on whether the alignment shall be \emph{global}, \emph{local}, or \emph{semi-global}.
%Also note that for linear gap penalties, where $G_0 = G_e = g$ the values of $E(i,j)$ and $F(i,j)$ are always determined by \autoref{eq:lingap} and don't need to be found by dynamic programming.

\paragraph{Local Alignments}
These compute an alignment with the highest score starting at any position $(q_i,s_j)$ and ending at any other position $(q_k,s_l)$ with $i \leq k \leq m, j \leq l \leq n$.
The parameter $\nu$ in \autoref{eq:H} is set to~$0$.
The \ac{DP} matrices are initialized as follows:
%by $H(0,0) = H(i,0) = H(0,j) = 0$, $E(0,0) = F(0,0) = F(i,0) = E(0,j) = -\infty$, $E(i,0) = -\alpha -(i-1) \cdot \beta$, and $F(0,j) = -\alpha - (j-1) \cdot \beta$.
%according to \eqref{eq:init_loc}.
\begin{align*} %\label{eq:init_loc}
    H(0,0) &= 0         & H(i,0) &= 0                       & H(0,j) &= 0 \notag \\
    E(0,0) &= -\infty   & E(i,0) &= -\alpha - \beta (i-1)   & E(0,j) &= -\infty \\
    F(0,0) &= -\infty   & F(i,0) &= -\infty                 & F(0,j) &= -\alpha - \beta (j-1)
\end{align*}

\paragraph{Global Alignments}
These always start at position~$(0,0)$ and end at position~$(m,n)$. Hence, their optimal score is always stored in cell $H(m,n)$.
The variable $\nu$ in \autoref{eq:H} is set to $-\infty$.
%\ac{DP} matrices are initialized by $H(0,0) = 0$, $E(0,0) = F(0,0) = F(i,0) = E(0,j) = -\infty$, $H(i,0) = E(i,0) = -\alpha - (i-1) \cdot \beta$, and $H(0,j) = F(0,j) = -\alpha - (j-1) \cdot \beta$.
The \ac{DP} matrices are initialized as follows:
\begin{align*} %\label{eq:init_glob}
    H(0,0) &= 0         & H(i,0) &= -\alpha - \beta (i-1)    & H(0,j) &= -\alpha - \beta (j-1) \\
    E(0,0) &= -\infty   & E(i,0) &= -\alpha - \beta (i-1)    & E(0,j) &= -\infty \\
    F(0,0) &= -\infty   & F(i,0) &= -\infty                  & F(0,j) &= -\alpha - \beta (j-1)
\end{align*}

\paragraph{Semi-global Alignments}
These do not penalize gaps at the beginning and the end. This leads to the same specifications as for local alignment but the optimal score is determined as the maximal value in the last row or column of $H$.

\paragraph{Traceback}
Score-only computations can be performed in linear space $\mathcal{O}(\min\{m, n\})$ and quadratic time $\mathcal{O}(m \cdot n)$.
%with respect to the length of the alignment targets.
Actual alignments producing this value can be determined by tracing back the predecessor information in the \ac{DP} matrices.
A divide-and-conquer approach~\cite{hirschberg75} consumes linear instead of quadratic space by recursively determining optimal midpoints of the \ac{DP} matrix---at the cost of doubling the amount of computed \ac{DP} cells in the worst case.

%Each cell of $H$ in \autoref{eq:H} depends on three neighboring cells (see \autoref{fig:dpmatrix}) which means that relaxing them in parallel can be done along minor diagonals. When relaxing all cells in a submatrix row of $H$, only the subproblem scores for the row above and one column left of the current cell are needed. If we want to compute submatrices in parallel, the first row and first column of a submatrix must have been computed earlier and the last row and last column must be kept available for the computation of subsequent submatrices (to the right of and below the current one).

%------------------------------------------------------------------------------
\subsection{GPU Computing}\label{}
%------------------------------------------------------------------------------
We briefly highlight a number of relevant features for GPU computing.

%using NVIDIA terminology.
\paragraph{NVIDIA GPUs}
These execute kernels with a number of independent thread blocks.
Each thread block is mapped onto exactly one \emph{streaming multiprocessor (SM)} and consists of a number of \emph{warps}.
Each warp contains 32 threads that are executed in lockstep---similar to vector instructions of SIMD units on CPUs.
GPUs contain several types of memory:
\begin{itemize}
    \item large but high latency off-chip global memory and
    \item fast but small on-chip shared and constant memory.
\end{itemize}
Nevertheless, the fastest way to access data is through usage of the thread-local register file.

\paragraph{AMD GPUs}
These GPUs exhibit a similar general structure.
They schedule kernels as a collection of independent \emph{workgroups} onto the \emph{workgroup processors (WGP)} of the GPU.
Each workgroup contains multiple \emph{wavefronts}.
A wavefront is AMD's counterpart to a warp.
This means that all \emph{work-items} within a wavefront are also executed in lockstep.
The most notable difference to NVIDIA GPUs is the number of threads executed in lockstep:
Starting with RDNA, wavefronts can be 32 or 64 threads wide, while previous architectures had a fixed wavefront width of 64.

%Modern GPUs provide instructions for warp-level collectives~\cite{cooperativegroups} in order to efficiently support communication of data stored in registers (such as shuffles or ballots) between threads within a warp without the need for accessing global or shared memory.
\paragraph{Warp Shuffles}
These operations are crucial features of our approach.
They allow for low latency communication and minimization of memory traffic.
In particular, we take advantage of the warp-level collectives \texttt{\_\_shfl\_down\_sync()} and \texttt{\_\_shfl\_up\_sync()}.
For example, the intra-warp communication operation
\begin{lstlisting}
R1 = __shfl_up_sync(0xFFFFFFFF, R0, 1, 32);
\end{lstlisting}
moves the contents of register \texttt{R0} in thread~$i$ within each warp to register \texttt{R1} in thread $i+1$ for $0 \le i < 31$.

%------------------------------------------------------------------------------
\subsection{The AnyDSL Compiler Framework}
\label{sec:anydsl}
%------------------------------------------------------------------------------

\anyseq is implemented in AnyDSL~\cite{DBLP:journals/pacmpl/LeissaBHPMSMS18}.
AnyDSL is a framework for staging domain-specific languages using partial evaluation (PE).
AnyDSL provides several benefits for the implementation of high-performance domain-specific languages:

\paragraph{Implementing DSLs without Writing a Compiler}
The programmer implements the domain-specific language as higher-order library functions in the AnyDSL host language \impala.
The DSL program is essentially a nested function call to these library functions.
By means of partial evaluation, the overhead of these nested calls is entirely removed:
The result is highly-optimized code that looks as if a programmer had manually specialized all these functions.
Technically, this a combination of the first Futamura projection~\cite{DBLP:conf/rims/Futamura82} and a so-called \emph{tagless interpreter}~\cite{DBLP:conf/aplas/CaretteKS07}.

Implementing DSLs this way typically promotes a more functional style of programming which allows for abstraction from concerns such as hardware dependence by separating these issue into dedicated functions that are passed to the core algorithms as high-order arguments.
While functional programming is often associated with slower execution, AnyDSL has demonstrated that the potential overhead of functional programming is succinctly eliminated using partial evaluation~\cite{DBLP:journals/tog/Perard-GayotMLH19, DBLP:conf/ipps/MullerS0MLKH20}.

\paragraph{Compilation to Heterogeneous Hardware}
The AnyDSL compiler infrastructure provides code generators for a variety of heterogeneous hardware targets: multi-core CPUs with vector instruction sets, NVIDIA and AMD GPUs, as well as different kinds of FPGAs.
AnyDSL exposes code generation to accelerators via compiler-known, higher-order functions that the programmer can use to map a certain piece of code to, say, a GPU.
This way, hardware-independent code is reusable for different targets and will be later on specialized with hardware-dependent code.
Therefore, when implementing a domain-specific library in AnyDSL, the implementer can readily choose among these code generators and is \emph{not} bothered with implementing their own code generators.
In turn, every DSL implemented in AnyDSL directly benefits from potential improvements of these code generators in the AnyDSL compiler.

\paragraph{Partial Evaluation}
\label{sec:anydsl:pe}
AnyDSL hinges strongly on partial evaluation and \anyseq makes use of partial evaluation in the way described above but also in several other ways that are described in \autoref{sec:kernelvariants} to elegantly optimize code even further.
Therefore, we briefly introduce the most important features of AnyDSL's partial evaluator.
Programmers control the partial evaluator via so-called \emph{filters}.
These are Boolean expressions of the form \lst|@(expr)| that annotate function signatures.
Each call site instantiates the callee's filter with the corresponding argument list.
If the expression evaluates to \texttt{true}, the call is \emph{guaranteed} to be specialized.
Additionally, the expression \lst|?expr| yields \texttt{true}, if \lst|expr| is known at compile time;
the expression \lst|$\texttt\textdollar$expr| is never considered constant by the evaluator.
For example, the following
\lst$@(?n)$ filter will only
specialize calls to \lst[language=impala]$pow$
if \lst[language=impala]$n$ is statically known at compile time:
\begin{lstlisting}[language=impala]
fn @(?n) pow(x: int, n: int) -> int { /*...*/ }
\end{lstlisting}
Thus, the call \lst|pow(x, 5)| produces a loop-less sequence of multiplications,
\lst|pow(3, 5)| evaluates to \lst|243| while
\lst|pow(x, $\texttt\textdollar$5)| remains untouched.
The fact that the specialization behavior is determined for each call site independently is called \emph{polyvariance}.
As syntactic sugar, \lst|@| is available as shorthand for \texttt{@(true)}.
This causes the partial evaluator to always specialize the annotated function.
Finally, \impala automatically annotates higher-order parameters for specialization.

%==============================================================================
\section{Methods}\label{methods}
%==============================================================================

We base our parallelization scheme on computing an independent alignment per warp, a group of synchronized threads that are executed in lockstep and communicate via collective operations such as shuffles or votes.
Threads in a warp thus can compute DP matrix cell values in a cooperative fashion.
In order to unlock the full potential of modern GPUs for alignment, we apply the following techniques:
\begin{itemize}
    \item Full in-register computation of the DP recurrence relations.
    \item Low latency communication of neighboring DP cells between threads through warp shuffles.
    \item Reduction of frequent sequence character loading from memory by employing an intra-warp communication scheme based on warp shuffles.
    \item Efficient scheme for substitution table lookups $\sigma(q_i,s_j)$.
    \item \ac{DP} matrix partitioning scheme for long sequences.
    \item Optional traceback in linear space.
\end{itemize}

%------------------------------------------------------------------------------
\subsection{Mapping Matrix Cells and Sequences to Threads}
\label{sec:mapping}
\begin{figure*}[t]
    \centerline{\includegraphics[width=1.0\textwidth]{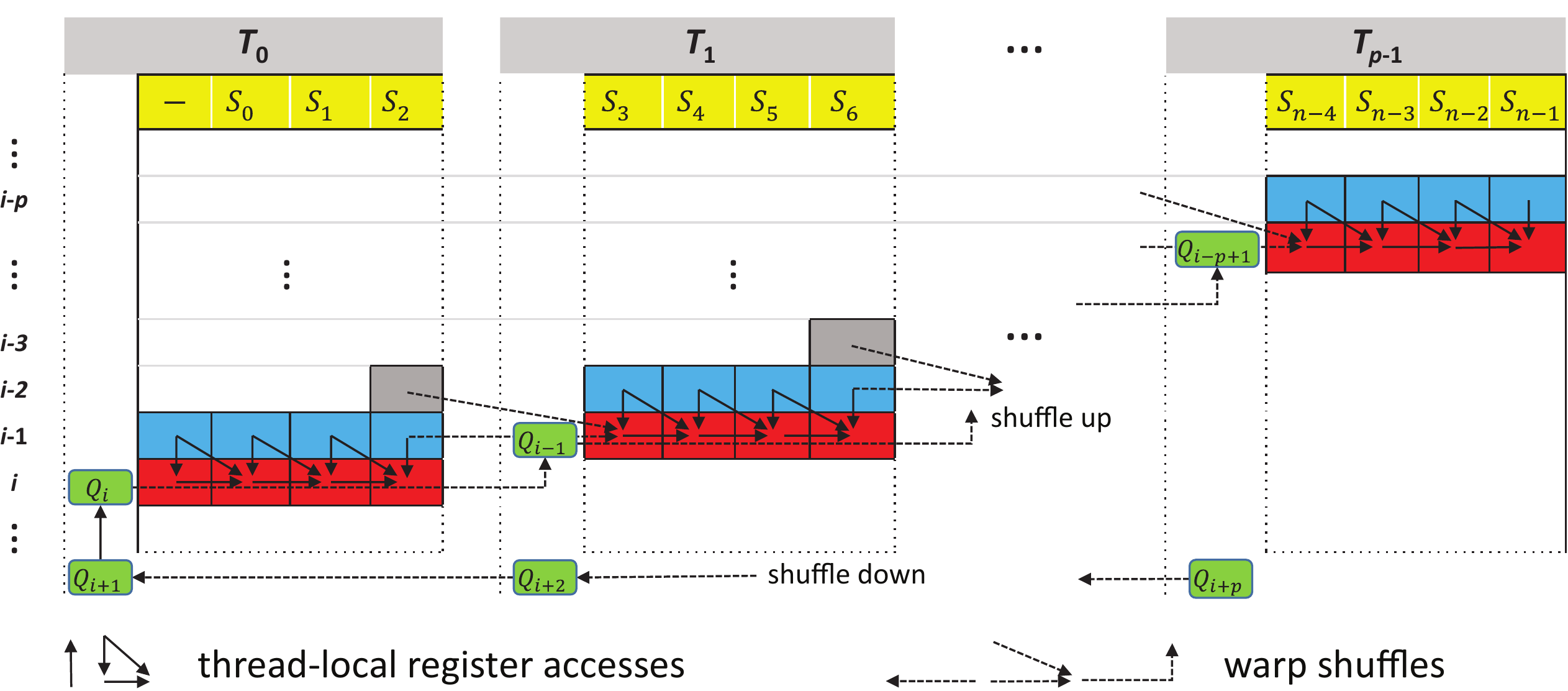}}
    \caption{
        Example of mapping a scoring matrix computation for comparing two (sub)sequences $Q$ and $S$ of length $n = k \cdot p - 1$ to a group of $p$ threads.
        Each thread computes $k$ matrix cells (red) in each iteration $i$ using the values of the antecedent row (blue) and an additional value from the second antecedent row (gray).
        Note that the characters of $S$ (yellow) are not accessed directly, instead a scoring profile is used to look up the score for pairwise comparisons.
        Threads communicate values in the rightmost of their $k$ columns as well as characters of $Q$ (green) by warp shuffle operations (dashed arrows).
    }
    \label{fig:mapping}
\end{figure*}

%------------------------------------------------------------------------------
Each warp in a thread block (a group of $p$ threads $T_0,...,T_{p-1}$ where $p=32$ for NVIDIA, $p=64$ for AMD) executed in lockstep, computes the score for aligning a (query) sequence $Q$ to a (subject) sequence $S$.
Assume $Q$ of length $m$ and $S$ of length $n=k \cdot p - 1$.
We use \emph{register tiling} to assign $k$ columns of the DP matrix to be calculated by each thread (see \autoref{fig:mapping}).
Computation proceeds along a wavefront in $m+p$ iterations.
In iteration $i$, thread $T_t$ computes $k$ adjacent cells of the DP matrix row $i-t$.
The maximum possible value of $k$ depends on the number of available registers and used data type.

According to the recurrence relations \autoref{eq:H}--\ref{eq:F} computation of each DP matrix cell depends on the values of its left, upper and upper-left neighbor (see  \autoref{fig:dpmatrix}).
All cells of the current and previous iteration are stored in thread-local registers.
Thread $T_t$ obtains the rightmost value of thread $T_{t-1}$ computed in the previous iteration by using a low-latency warp shuffle-up instruction.

At the beginning each thread $T_t, 0 \leq t < p$ loads $k$ characters of the subject sequence $S$ from global memory.
% $S_j, t \cdot k-1 \leq j < t \cdot k + k - 1$.
Note that the characters of $S$ remain the same for each column of the DP matrix while the required characters from $Q$ vary.
The loaded subject characters are used to create a \emph{scoring profile} in shared memory; i.e., a lookup table that stores the resulting substitution score $\sigma$ of all pairwise comparisons of each subject character with all characters of the alphabet ${A,C,G,T}$.
In subsequent steps, this profile is used to obtain scores for the pairwise comparisons of subject and query sequence characters.
This approach avoids branch divergence associated with handling cases of mismatching character pairs differently from matching ones.
Furthermore, a 4-character-wide data type allows for looking up 4 neighboring cells in the substitution table with a single shared memory access to the scoring profile.
Note that creating the scoring profile requires only one linear pass over the input sequence which results in a negligible runtime compared to the subsequent DP matrix computation which is quadratic in the input sequence length.

In order to avoid repeated, expensive reading from global memory, we only load new query characters every $p$ iterations and otherwise exchange them between threads via warp shuffles with two registers $c_{q0}$ and $c_{q1}$ per thread.
Register $c_{q0}$ stores the query character $Q_{i-t-1}$ needed for the current computation by thread $t$ in iteration $i$ while register $c_{q1}$ stores the character needed for the next computation.
At the start of each iteration, thread $0$ copies the current value of $c_{q1}$ to $c_{q0}$.
Afterwards, a warp shuffle-up and a warp shuffle-down update $c_{q0}$ and $c_{q1}$, respectively, with characters from neighboring threads.
Only in iterations $0 \leq i < m + p$ with $i \mod p = 0$ each thread $t$ loads a new query character $Q_{i+t}$ from memory and stores it in $c_{q1}$.

To compute the current matrix row, each thread first looks up the $k$ substitution scores $\sigma(c_{q0},S_{x})$ for $x = t \cdot k+j-1$ for $x \geq 0, 0 \leq j \leq k$ from the scoring profile and subsequently computes the recurrence relations \autoref{eq:H}--\ref{eq:F}.
This involves several maximum, addition, and subtraction operations.

On modern GPUs from both NVIDIA and AMD the maximum number of registers per SIMD unit is limited to 255 and 256, respectively.
This restricts the maximum number of columns $k$ that are computable by one thread.
In order to overcome this limitation, we partition the workload into $l$ non-overlapping submatrices of size $(m+1) \times \frac{n+1}{k \cdot p}$ that are computed by a thread group in $l$ stages from left to right.
DP cells in the right column of thread $T_{p-1}$ are stored in shared memory or global memory (depending on the sequence length) and loaded by thread $T_0$ in the next stage.
In order to avoid repeated, expensive accesses to global memory, these values are only loaded/written every $p$ iterations and otherwise exchanged between threads via warp shuffles in a similar way to reading query sequence characters as explained above.

\autoref{lst:kernel} shows Impala pseudo-code of our kernel that computes global alignments with a linear gap scoring scheme using a single warp and a single stage, i.e., for a subject sequence that has at maximum $p \cdot k$ characters. \lst|update_matrix()| computes $k$ DP cells per thread (red cells in \autoref{fig:mapping}) using thread-local in-register computation and shared memory lookups to the scoring profile. Required communication to access neighboring DP cells (\lst|shfl_up_sync(H.read(k),1)|) or neighboring query sequence characters (\lst|shfl_up_sync(c_q0,1)|, \lst|shfl_down_sync(c_q1,1)|) occur outside this function using low latency warp shuffles (dashed arrows in \autoref{fig:mapping}). Within the for-loop, global memory is only accessed every $p$ iterations (\lst|if i%p == 0 {...}|) to load new query characters (green values in \autoref{fig:mapping}) in a coalesced fashion that are stored in thread-local registers (variable \lst{c_q1}).
Prior to the for-loop \lst|init_profile()| initializes scoring profiles from subject sequences in shared memory.

\lstset{language=impala}
\begin{lstlisting}[label=lst:kernel,caption={Pseudo-code for a single-stage linear penalty scoring kernel illustrating the basic data flow pattern of our algorithm.},captionpos=b,float]
// @-annotation instructs the partial evaluator to
// specialize this function wrt parameters p and k.
fn @(?p & ?k) scoring (
    p: i32,         // #threads in a group
    k: i32,         // #matrix columns per thread
    grid: WorkItm,  // block and thread indices
    query: SequenceView,
    subjects: SequenceBatchView,
    scores: ScoresView  // results
) -> () {
    // register storage:
    let H = make_unrolled_array(k); // k matrix cells
    let mut H_left: Score;          // from left neighbor
    let mut H_diag: Score;          // -"-
    let mut c_q0: Char;             // current query char
    let mut c_q1: Char;             // next query char
    // subject -> scoring profile in shared memory
    let P = make_scoring_profile(4*p*k);
    init_profile(P, S, grid);
    // initialize DP cells
    init_matrix(H, H_left, H_diag, p, k, grid);
    // load one query value per thread
    c_q1 = query.read(grid.tidx%p);
    c_q0 = if grid.tidx % p == 0 { c_q1 } else { -INF };
    // wavefront loop
    for i in range(1, m + p) {
        // compute k DP cells per thread
        // using registers and score profile only
        update_matrix(H, H_left, H_diag, P, c_q0, k);
        H_diag = H_left;
        // rightmost cell to neighboring thread
        H_left = shfl_up_sync(H.read(k), 1);
        if grid.tidx % p == 0 { H_left = -INF; }
        // load new query char every p iterations
        if i%p == 0 { c_q1 = query.read(i+grid.tidx%p); }
        // shuffle query registers
        c_q0 = shfl_up_sync(c_q0, 1);
        if grid.tidx % p == 0 { c_q0 = c_q1; }
        c_q1 = shfl_down_sync(c_q1, 1);
    }
    output_score(H.read(k), scores, grid, p, k);
}
\end{lstlisting}

%------------------------------------------------------------------------------
\subsection{GPU Kernel Variants}
\label{sec:kernelvariants}
%------------------------------------------------------------------------------

We employed AnyDSL's user-guided partial evaluation to generate GPU kernels that are tailored to different hardware requirements, alignment schemes, maximum input sequence lengths as well as scoring schemes with their associated penalty values, which are often known at compile time.

Data views are used to abstract the memory access to different parts of the input sequence batch, scoring profile storage, storage for intermediate scores and storage for the final scores.
All reading and writing data accesses are hidden behind accessor functions whose calling overhead is fully eliminated by PE. Furthermore, since functions are first-class citizens in Impala, one can easily change data access schemes by assigning different functions to members of ``data view'' structs.
This allows us to implement code in a data-source-agnostic way and also enables easy exploration of different data layouts during development.
We obtain optimized kernels for the alignment of sequences with a known maximum length by setting this value in the corresponding data view and feeding it to the partial evaluator.

In a similar manner, we specialize most parameters of the alignment and scoring scheme by selecting appropriate functions that provide the desired behavior at compile time.
Such functions are grouped in structs and are used to determine the substitution function $\sigma$, the strategy for initializing cells in the leftmost column and topmost row of the DP matrix, and the lookup and tracing of the optimum score.
Scoring strategies for linear and affine gap penalties are factored out to separate kernels as the need to perform additional computations on matrices $E$ and $F$ leads to a significantly altered code structure throughout a scoring function.
Partial evaluation combines these kernels with the desired scoring function without performance overhead.

In addition to the alignment score we also implemented the full traceback of the actual alignment.
For sequences with a total length of up to 128 characters we explicitly store the optimum cell predecessors which leads to quadratic memory overhead.
We use \citeauthor{hirschberg75}'s traceback algorithm~\cite{hirschberg75} for longer sequences as it only requires linear memory overhead at the expense of doubling the amount of total score computations.

Maximum performance is only achieved if the arrays for the matrix cell values associated with a thread are all stored in registers.
Unlike objects of fundamental or aggregate type, arrays defined in a CUDA kernel's scope are not always guaranteed to be stored in registers and may be put in much slower constant memory instead.
The compilers by NVIDIA or AMD usually decide this based on heuristics without providing guarantees to the developer.
However, using AnyDSL we defined an \emph{unrolled array} type that is usable like an ordinary array while its elements are actually put into scalar registers---combining the convenience of index-based access with the performance of register variables.
The size of an unrolled array is fixed at compile time and access to elements is done by mapping array indices to variables using recursive conditions that are partially evaluated at compile time.

Score computation requires several smaller loops, e.g., over all 4 possible DNA characters.
According to our observations such loops often benefit from loop unrolling.
Using Impala's for-loop abstraction, we simply exchange a regular range-based loop iterator with an unrolling one.
Unlike traditional compiler pragmas which are non-binding requests, AnyDSL \emph{guarantees} that unrolling takes place when desired, because the \lstinline|unroll| iterator is implemented using partial evaluation filters (see \autoref{sec:anydsl:pe}).
This allows for easy, step-by-step tuning of kernels guided by reliable performance experiments.

All score computations exclusively use floating-point arithmetic since we observed that it consistently leads to higher throughput on all our test systems compared to integer arithmetic.
In addition, we also employed half-precision arithmetic instructions supported by NVIDIA A100 and AMD MI100 accelerators that allow for the simultaneous execution of one operation on two half-precision values.
Besides instructions for packing (extracting) 2 single-precision values into (from) a pair of half-precision values, we used addition and maximum\footnote{The required \lst|half2| max comparison function is supported on A100 and MI100 but not on the utilized GV100 and RTX3090 GPUs.} operations that operate on pairs of half-precision values to compute the DP recurrence relations.
We compute two independent alignments at the same time in the lower and the upper part of a packed pair of half-precision values thus effectively doubling the number of cells computed per thread.

Note that there are no rounding issues since the actual values of the computation are always integer.
Furthermore, the limited range of half-precision FP numbers is also not an issue since in practice the absolute result of $\sigma$ or the gap penalties values is usually well below~10.\footnote{One could only exceed the range of representable integers during score computation if one encountered a stretch of only mismatches or only matches that is longer than the $p \cdot k$ columns that are computed in one stage of the algorithm.}

Partial evaluation filters (see function signature in \autoref{lst:kernel}) were used to specialize GPU kernels for different values of $p$ and $k$ based on the available hardware features.

For each target platform the Impala runtime library as well as our code base ships a separate file in which hardware specific aspects are mapped to Impala types and functions.
This provides a uniform, architecture-agnostic interface that includes abstractions for memory allocation, kernel invocation and execution of intrinsics like warp shuffles and half-precision arithmetic operations.
This allows for development across GPU platforms without code duplication as the same code can be compiled for different GPU architectures by selecting the appropriate mapping files at compile time.

%==============================================================================
\section{Evaluation}\label{performance}
%==============================================================================

%------------------------------------------------------------------------------
\subsection{Experimental Setup}

We compared \anyseq to three GPU alignment libraries: NVBIO\,1.1, GASAL\,2 and ADEPT, to the established CPU sequence alignment library SeqAn\,2.4.0, and to AnySeq\,1 executed on both CPU and GPU.

We evaluated the experiments on the following systems:
\begin{enumerate}
% Monster 2
\item Dual Intel Xeon Gold 6130 (Skylake) CPUs, 192\,GB of DDR4 RAM, 2 NVIDIA GV100.
% alte Konfiguration von Monster 1
\item Dual Intel Xeon E5-2683, 256\,GB RAM, NVIDIA RTX3090.
% DGX-A100
\item NVIDIA DGX-A100 with 8 A100 GPUs, Dual AMD Rome 7741, 2\,TB RAM.
% MI 100
\item Dual Intel Xeon Gold 6130, 128 GB RAM, AMD MI100.
\end{enumerate}

CPU-based alignment experiments with AnySeq\,1 and SeqAn were run on System~1 using 32~threads.
GPU experiments were conducted on one of the GPUs in each of the systems.
\anyseq was compiled with the AnyDSL version from March 2021 which is based on LLVM~10.0.
The GPU alignment libraries GASAL\,2, ADEPT and the relevant parts of NVBIO were compiled with nvcc~10.2 with gcc~9.3.0 as host compiler for CUDA-enabled GPUs and compiled for use with the MI100 GPU using the HIPIFY tool that is part of AMD's ROCm 4.3.1 developer tool suite.
Benchmark programs that use SeqAn were compiled with g++~9.3.0.

We computed global, semi-global and local alignments using both a simple scoring scheme with $+2$ for a match, $-1$ for a mismatch and a linear gap penalty of $1$ and an affine scoring scheme using the same match and mismatch penalties and gap penalties $ \alpha=2$ and $\beta=1$.
AnySeq\,1, SeqAn, NVBIO and GASAL offer optimized implementations for all combinations of alignment types and gap penalty schemes while ADEPT is primarily optimized for local alignments using an affine gap penalty scheme.

We evaluated the alignment speed for two common tasks that are major runtime contributors in many bioinformatics pipelines:
\begin{itemize}
    \item pairwise alignment of short DNA snippets (called sequencing reads), and
    \item alignment of long sequencing reads against a longer genomic sequence.
\end{itemize}
Speeds are reported by measuring runtimes (excluding PCIe data transfers) and converting them into the number of DP matrix cell updates that are performed per second in GCUPS or TCUPS, i.e., billions (giga) or trillions (tera) of cell updates per second.

All read data sets were created from human chromosome 10 sequence GRCH38\_chr10 using simulator tools. We generated a total of six sets each consisting of 3536 Illumina short to medium-size reads with lengths of 125, 250, 512, 1024, 2048 and 4096 characters using Mason\,2.0.9 with default settings.
For each combination of short read data set, system, alignment library and alignment scheme the resulting alignment speed was obtained as the median speed of 10 runs (arithmetic mean of the 5th and 6th decile value) where each run computed all $3536^2\approx 12.5$ million pairwise alignments of the reads in a set.

As long read sequencing technologies have been gaining importance in recent years, we also included a set of PacBio long reads simulated using PBSIM\,1.0.3.
Individual read lengths in our Pacbio read set vary from 4,442 to 57,571 characters with an average length of 21,223 and a standard deviation of 6,482.
% load imbalances produced by varying read lengths
The Pacbio reads were aligned against the original genomic sequence from which they were created.
As for the short reads, each speed result reported is the median speed of 10 runs.

Detailed information about the usage of the two simulators (MASON and PBSIM) and the constructed datasets are also provided on the \anyseq repository (
\url{https://github.com/AnyDSL/anyseq}).

Note that due to the quadratic time complexity, PCIe transfers are typically negligible; e.g., for the 512bp Illumina dataset used in our tests, data transfers only take 0.36 ms. That compares to a computation time of around 866 ms for global alignment with linear gaps on an MI100.
Furthermore, PCIe data transfers can be overlapped with computation using batching. Similar values hold for the transfer of the results back to the CPU (note that in many scenarios this is not even required since the results could be further processed on the GPU).

%------------------------------------------------------------------------------
\subsection{Efficiency Analysis}

%------------------------------------------------------------------------------
% altes AnySeq habe ich nie auf A100 oder MI100 ausgeführt nur auf Titan V, GV100 und der RTX3090
% die AnySeq CPU-Werte habe ich auch zu AnySeq1 verschoben, sonst währe das glaube ich verwirrend
\begin{table*}[t]
\centering
\begin{tabular}{cl|rrrrr|rrrrr}
    \toprule
      &            & \multicolumn{5}{c|}{Score Only}              &     \multicolumn{5}{c}{Traceback}    \\
      & Library    & AVX512 &   GV100 & RTX3090 &  A100 &   MI100 &      AVX512 &   GV100 & RTX3090 &  A100 &  MI100 \\
    \midrule          % score comp --------- & traceback --\\
    \multirow{4}{*}{\rotatebox[origin=c]{90}{125\,bp}}
      & \anyseq    &     -- &    \bf{1142} &    \bf{1877} &  \bf{2459} &    \bf{2909} &          -- &     \bf{641} &     \bf{997} &  \bf{1282} &  \bf{1461}  \\
      & AnySeq\,1  &    208 &     261 &     382 &    -- &      -- &         \bf{148} &     177 &     242 &    -- &    --  \\
      & GASAL\,2   &     -- &     126 &     198 &   233 &     289 &          -- &      92 &     147 &   164 &   198  \\
      & NVBIO      &     -- &     134 &     209 &   228 &     278 &          -- &      25 &      31 &    39 &    47  \\
      & SeqAn      &    \bf{219} &      -- &      -- &    -- &      -- &         133 &      -- &      -- &    -- &    --  \\
    \midrule          % score comp --------- & traceback --\\
    \multirow{4}{*}{\rotatebox[origin=c]{90}{512\,bp}}
      & \anyseq    &     -- &    \bf{1712} &    \bf{2775} &  \bf{3268} &    \bf{3754} &          -- &    \bf{1012} &    \bf{1613} &  \bf{1892} &  \bf{2123}  \\
      & AnySeq\,1  &    206 &     241 &     336 &    -- &      -- &         125 &     164 &     219 &    -- &    --  \\
      & GASAL\,2   &     -- &      48 &      82 &    97 &     116 &          -- &      35 &      60 &    71 &    84  \\
      & NVBIO      &     -- &      39 &      54 &    92 &     107 &          -- &      21 &      32 &    39 &    46  \\
      & SeqAn      &    \bf{214} &      -- &      -- &    -- &      -- &         \bf{134} &      -- &      -- &    -- &    --  \\
    \midrule          % score comp --------- & traceback --\\
    \multirow{3}{*}{\rotatebox[origin=c]{90}{Pacbio}}
      & \anyseq    &     -- &    \bf{1297} &    \bf{1719} &  \bf{2310} &    \bf{2511} &          -- &     \bf{578} &     \bf{971} &  \bf{1138} &  \bf{1213}  \\
      & AnySeq\,1  &    192 &     212 &     296 &    -- &      -- &         111 &     147 &     206 &    -- &    --  \\
      & GASAL\,2   &     -- &      40 &      61 &    76 &      89 &          -- &      28 &      51 &    59 &    69  \\
      & SeqAn      &    \bf{208} &      -- &      -- &    -- &      -- &        \bf{124} &      -- &      -- &    -- &    --  \\
    \bottomrule
\end{tabular}
\caption{
    Median speeds in GCUPS for global alignments using a linear gap penalty scheme for a) 12.5 million pairwise alignments of 125\,bp short reads, b) 12.5 million pairwise alignments of 512\,bp short reads and c) alignments of long Pacbio reads ($21223\pm6482$\,bp) against a long genomic sequence ($\approx135$ million bp). CPU experiments were run on System~1. Best values per column are indicated in bold.
}
\label{tab:perf_top}
\end{table*}

\begin{table}[t]
  \centering
  \caption{Efficiency of \anyseq for global alignment with affine gap penalties on different GPUs. Achieved TCUPS is the maximal performance of \anyseq in \autoref{fig:perf_scaling_global_affine}.
  (TPP = theoretical peak performance.)}
\begin{tabular}{ l | c| c| c | c }
    \hline
    &  V100 & RTX3090 & A100 & MI100  \\
    \hline
    \hline
    FP32 cores & 5120 & 10496 & 6912 & 7680 \\
    Boost Clock (GHz) & 1.91 & 1.70 & 1.41 & 1.50 \\
    \hline
    Cycles Cell Update & 7 & 7 & 3.5 & 3.5 \\
    TPP in TCUPS & 1.40 & 2.55 & 2.78 & 3.29 \\
    Achieved TCUPS & 1.33 & 2.09 & 2.49 & 2.82 \\
        \hline
    Efficiency & 95\% & 82\% & 89\% & 86\%\\
    \hline
    \hline
\end{tabular}
\label{tab:efficiency}
\end{table}

We first analyze if our approach is able to effectively remove overheads occurred by memory accesses by modeling the  {\it theoretical peak performance} (TPP) in terms of cell updates per second (CUPS) of the utilized GPU hardware as:
\begin{equation}
\mathrm{TPP} = \frac{\mathrm{\#Cores}  \times \mathrm{Clock}} {\mathrm{Cycles\_per\_cell\_update}}
%= \frac{2 \times 5120  \times 1.85 \text{ GHz}} {4} = 4.7 \text{ TCUPS}
\label{eq:peak}
\end{equation}
Cycles\_per\_cell\_update in \autoref{eq:peak} models the maximum attainable performance constrained by the algorithm structure.
In our case it refers to the minimal number of clock cycles needed by an individual FP32-core of the utilized GPU to calculate one DP matrix cell.

We use global alignment with affine gap penalties with Illumina reads as case study to analyze to what extend \anyseq makes optimal use of the hardware. The number of required arithmetic operations in our implementation to calculate a single \ac{DP} cell is 7 (i.e. 4 maximum instructions  and 3 additions/subtractions). These values are determined by \autoref{eq:H} (1 add, 2 max)\footnote{Only 2 maximum instructions are required in \autoref{eq:H} since $\nu$ is $-\infty$ for global alignments.} and optimized versions of \autoref{eq:E} and \ref{eq:F} (2 sub, 2 max)\footnote{Subtractions in \autoref{eq:E} and \ref{eq:F} can be reduced from 4 to 2 since $\beta$ only needs to be subtracted once (using $\max\{ E(i-1,j), F(i,j-1) \} - \beta$) and $\alpha$ also needs to be subtracted only once from each value in $H$.}. This leads to 7 required clock cycles per cell update (when using single-precision arithmetic on GV100 and RTX3090) and 3.5 required clock cycles (when using \lst|half2| arithmetic on A100 and MI100). Note that we disregard the lookup operation $\sigma(q_i,s_j)$ since it can in principle be performed concurrently to computation. \autoref{tab:efficiency} shows that \anyseq is able to achieve efficiencies of over 80\% on all utilized GPUs, which shows that our approach is able to effectively minimize overheads from memory accesses.

\begin{figure*}[p]
    \input{bars.tex}
    \label{fig:perf_overview}
\end{figure*}

\pgfplotscreateplotcyclelist{amlist}{
    blue, every mark/.append style={fill=.!80!black},mark=*\\
    red, every mark/.append style={fill=.!80!black},mark=square*\\
    black, every mark/.append style={fill=.!80!black},mark=triangle*\\
    densely dashed, blue, every mark/.append style={fill=.!80!black},mark=*\\
    densely dashed, red, every mark/.append style={fill=.!80!black},mark=square*\\
    densely dashed, brown, every mark/.append style={fill=.!80!black},mark=triangle*\\
}
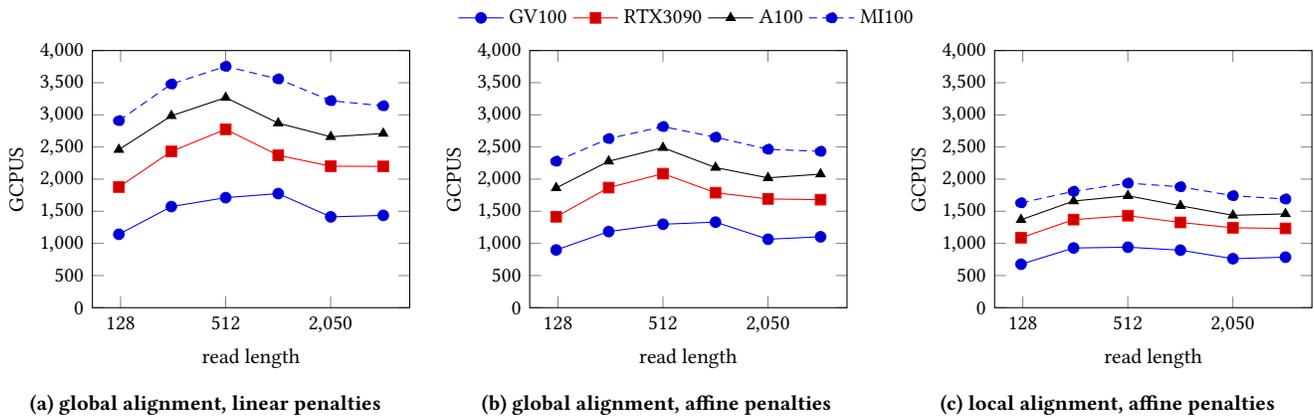
\begin{figure*}[p]
\begin{subfigure}[b]{5.8cm}
    \centering
    \begin{tikzpicture}
        \tikzstyle{every node}=[font=\small]
        \begin{axis}[
                width=\columnwidth,
                cycle list name = amlist,
                xmode=log,log basis x={2},
                log ticks with fixed point,
                xlabel={read length}, xlabel style={yshift=0},
                ylabel={GCPUS}, ylabel style={yshift=0},
                legend pos=outer north east,
                legend style={font=\tiny},
                ymin=0, ymax=4000,
                ytick={0,500,...,5000},
            ]
            \addplot coordinates { (125, 1142) (250, 1573) (512, 1712) (1024, 1775) (2048, 1413) (4096, 1436) };
            % \addlegendentry{GV100}
            \addplot coordinates { (125, 1877) (250, 2431) (512, 2775) (1024, 2371) (2048, 2201) (4096, 2198) };
            % \addlegendentry{RTX3090}
            \addplot coordinates { (125, 2459) (250, 2985) (512, 3268) (1024, 2871) (2048, 2659) (4096, 2710) };
            % \addlegendentry{A100}
            \addplot coordinates { (125, 2909) (250, 3479) (512, 3754) (1024, 3558) (2048, 3220) (4096, 3140) };
            % \addlegendentry{MI100}
        \end{axis}
    \end{tikzpicture}
    \caption{global alignment, linear penalties}
    \label{fig:perf_scaling_global_linear}
\end{subfigure}\hfill
\begin{subfigure}[b]{5.8cm}
    \centering
    \begin{tikzpicture}
        \tikzstyle{every node}=[font=\small]
        \begin{axis}[
                width=\columnwidth,
                cycle list name = amlist,
                xmode=log,log basis x={2},
                log ticks with fixed point,
                xlabel={read length}, xlabel style={yshift=0},
                ylabel={GCPUS}, ylabel style={yshift=0},
                %legend pos=outer north east,
                legend style={font=\tiny,at={(0.5,1.05)},anchor=south,legend columns=4,draw=none},
                ymin=0, ymax=4000, %ymax=3000,
                ytick={0,500,...,4000},
            ]
            \addplot coordinates { (125,  898) (250, 1184) (512, 1298) (1024, 1331) (2048, 1065) (4096, 1103) };
            \addlegendentry{GV100}
            \addplot coordinates { (125, 1414) (250, 1868) (512, 2085) (1024, 1789) (2048, 1692) (4096, 1680) };
            \addlegendentry{RTX3090}
            \addplot coordinates { (125, 1863) (250, 2278) (512, 2489) (1024, 2180) (2048, 2020) (4096, 2078) };
            \addlegendentry{A100}
            \addplot coordinates { (125, 2278) (250, 2630) (512, 2817) (1024, 2652) (2048, 2464) (4096, 2432) };
            \addlegendentry{MI100}
        \end{axis}
    \end{tikzpicture}
    \caption{global alignment, affine penalties}
    \label{fig:perf_scaling_global_affine}
\end{subfigure}\hfill
\begin{subfigure}[b]{5.8cm}
    \centering
    \begin{tikzpicture}
        \tikzstyle{every node}=[font=\small]
        \begin{axis}[
                width=\columnwidth,
                cycle list name = amlist,
                xmode=log,log basis x={2},
                log ticks with fixed point,
                xlabel={read length}, xlabel style={yshift=0},
                ylabel={GCPUS}, ylabel style={yshift=0},
                legend pos=outer north east,
                legend style={font=\tiny},
                ymin=0, ymax=4000, %ymax=2200,
                ytick={0,500,...,4000},
           ]
           \addplot coordinates { (125, 676) (250, 929) (512, 941) (1024, 895) (2048, 762) (4096, 786) };
           %\addlegendentry{GV100}
           \addplot coordinates { (125, 1088) (250, 1370) (512, 1431) (1024, 1326) (2048, 1243) (4096, 1233) };
           %\addlegendentry{RTX3090}
           \addplot coordinates { (125, 1368) (250, 1659) (512, 1741) (1024, 1586) (2048, 1437) (4096, 1459) };
           %\addlegendentry{A100}
           \addplot coordinates { (125, 1631) (250, 1810) (512, 1937) (1024, 1881) (2048, 1742) (4096, 1690) };
           %\addlegendentry{MI100}
        \end{axis}
    \end{tikzpicture}
    \caption{local alignment, affine penalties}
    \label{fig:perf_scaling_local_affine}
\end{subfigure}
% \begin{subfigure}{9cm}
%     \centering
%     \begin{tikzpicture}
%         \tikzstyle{every node}=[font=\small]
%         % \pgfplotsset{every axis legend/.append style={ at={(0.5,0.03)}, anchor=north}}
%         \begin{axis}[
%                 width=0.85\columnwidth,
%                 cycle list name = amlist,
%                 xmode=log,log basis x={2},
%                 log ticks with fixed point,
%                 xlabel={read length}, xlabel style={yshift=0},
%                 ylabel={GCPUS}, ylabel style={yshift=0},
%                 legend style={font=\tiny},
%                 legend pos=north east,
%                 ymin=0, ymax=320,
%                 ytick={0,50,...,400},
%             ]
%             \addplot coordinates { (125, 98) (250, 88) (512, 64) (1024, 32) };
%             \addlegendentry{GASAL\,2, GV100}
%             \addplot coordinates { (125, 144) (250, 120) (512, 72) (1024, 50) };
%             \addlegendentry{GASAL\,2, RTX3090}
%             \addplot coordinates { (125, 180) (250, 131) (512, 90) (1024, 61) };
%             \addlegendentry{GASAL\,2, A100}
%
%             \addplot coordinates { (125, 110) (250, 58) (512, 26) (1024, 11) };
%             \addlegendentry{NVBIO, GV100}
%             \addplot coordinates { (125, 153) (250, 78) (512, 38) (1024, 20) };
%             \addlegendentry{NVBIO, RTX3090}
%             \addplot coordinates { (125, 198) (250, 102) (512, 49) (1024, 27) };
%             \addlegendentry{NVBIO, A100}
%         \end{axis}
%     \end{tikzpicture}
%     \caption{}
%     \label{fig:perf_scaling_local_affine2}
% \end{subfigure}

    \caption{Median speed in GCPUS achieved for pairwise score computation using \anyseq for reads of varying lengths for a) global alignment, linear penalties; b) global alignment, affine penalties; c) local alignment, affine penalties.}
\end{figure*}

%------------------------------------------------------------------------------
\subsection{Performance Comparison}

\autoref{tab:perf_top} shows the median speeds in GCUPS  for computing global alignments using a linear gap penalty scheme with and without performing a full traceback for the different libraries (\anyseq, GASAL\,2, NVBIO, SeqAn, AnySeq\,1) and three different read sets (125bp, 512bp, PacBio).
%: pairwise alignment of 125\,bp reads, pairwise alignment of 512\,bp reads and alignment of long Pacbio reads to its longer origin sequence.
Computing only the global alignment score with a linear penalty scheme is the scenario for which all libraries deliver their maximum performance on all four test systems.
This is to be expected since no additional DP cells for affine gap penalties have to be maintained and the currently best local score does not need to be tracked.

\anyseq clearly outperforms all other tested GPU libraries in each test case.
When aligning short reads of length 125bp \anyseq achieves a minimum speedup of 4.4$\times$ and a median speedup of 9.3$\times$ over all other tested GPU libraries on the same hardware and a minimum speedup of 3.6$\times$ and a median speedup of 7.6$\times$ when also computing the traceback.
The difference becomes even more pronounced for reads of length 512bp where \anyseq outperforms the other GPU libraries by a minimum factor of 7.1 and a median factor of 34.5 for score computation using the same hardware setup and at least a factor of 6.2 and a median factor of 27.9 when also computing the traceback.
Again using the same hardware setup, the performance advantage over the other GPU libraries for aligning long Pacbio reads with a high length variance is of similar magnitude, where \anyseq is at least 5.8 times faster with a median speedup of 28.2$\times$ for score computation and at least 3.9$\times$ faster with a median speedup of 18.3$\times$ when performing a full traceback.
Summarizing, \anyseq provides a minimum speedup of at least 3.6$\times$ and a median speedup of 19.2$\times$ across all tested alignment scenarios and hardware setups compared to GASAL\,2, NVBIO and AnySeq\,1.

State-of-the-art CPU-based alignment libraries like SeqAn or AnySeq\,1 can outperform or closely match the performance of the fastest currently available GPU libraries for the same alignment task when running on a modern workstation CPU.
However, for the first time, \anyseq provides a clear benefit of using a costly GPU accelerator over a workstation CPU.
It outperforms both tested CPU libraries running with 32 threads and using AVX512 SIMD instructions by at least a factor of 4.3 with a medium speedup of 5.9$\times$ using the slowest tested GV100 GPU and a factor of at least 9.8 with a median speedup of 13.2$\times$ on an MI100 which was the fasted GPU in our benchmarks.

Aligning short reads against other short reads is one of the most prevalent tasks in bioinformatics pipelines.
\autoref{fig:perf_overview} gives an overview of speeds achieved for the pairwise alignment of 125\,bp Illumina short reads for different combinations of alignment schemes (global, semi-global, linear), gap penalty scheme (linear or affine) and result type (score only, traceback).
These results reveal that \anyseq's GPU implementation consistently outperforms all tested competitors.
The setup in which all libraries perform worst is computing the traceback of local alignments with affine gap penalties, yet \anyseq is able to achieve up to 757~GCUPS on an A100 GPU in this scenario while the fastest competitor GASAL\,2 achieves only 124~GCUPS on the same GPU and the fastest CPU implementation achieves only 82~GCUPS.

We performed several experiments with varying read lengths in order to investigate its influence on the resulting speed in batched alignment scenarios on the GPU.
On the newer RTX3090, A100 and MI100 GPUs \anyseq shows a continuous increase in speed up to a read length of 512\,bp for all score-only pairwise alignment scenarios including the fastest setup (global alignment, linear gaps) as well as the slowest setup (local alignment, affine gaps) as shown in \autoref{fig:perf_scaling_global_linear}--\subref{fig:perf_scaling_local_affine}.
Global alignments on the GV100 GPU are fastest for reads of length 1024\,bp while affine alignments are fastest for reads of length 256\,bp.
Across all tested setups, \anyseq aligns reads of up to 4096\,bp no more than 20\% slower than its best performance at 256 or 512\,bp. This shows the effectiveness of our partitioning scheme for different sequence lengths.
%In contrast, GASAL\,2 and NVBIO show a continuous decline in performance with increasing read length while being slower overall than AnySeq as can be seen in \autoref{fig:perf_scaling_local_affine2}.

%------------------------------------------------------------------------------
\section{Related Work}\label{relwork}
%------------------------------------------------------------------------------

Parallelization of pairwise sequence alignment has been investigated mainly in the context of two types of application scenarios:
\begin{enumerate}
    \item computing a single alignment of two for very long DNA sequences (typically whole genomes or chromosomes), and
    \item computing batches of different alignments of shorter length.
\end{enumerate}
Widely adopted parallelization schemes are \textit{inter-sequence} (computes \ac{DP} matrix cells of a number of independent alignment tasks in parallel), \textit{intra-sequence} (computes \ac{DP} matrix cells of a single alignment in parallel), and their hybrid combination.

\subsection{CPU Implementations}
Approaches on CPUs for processing \ac{NGS} data where a large number of alignments needs to be computed typically combine SIMD vectorization with multi-threading~\cite{rahn2018generic,misra2018performance,parasail,ssw}.
For the alignment of a single pair of long genomic sequences, most approaches employ wavefront parallelism for intra-sequence parallelization~\cite{rucci2018swifold,hou2016aalign,swaphils}. This approach takes advantage of the dependency relationship
in the recurrence relation: each cell depends on its left, upper, and upper-left neighbor.
An additional level of coarse-grained parallelism (e.g. on CPU/GPU clusters) has been proposed based on speculative execution or compensation-based parallelism~\cite{de2016cudalign,Maleki,HouWFVL18}.

\subsection{GPU Implementations}
Existing GPU-based approaches often compute an independent alignment per CUDA thread block whereby threads  compute  cells  along  a  minor  diagonal  of  the  \ac{DP}  matrix in  parallel. Data between neighboring diagonals is communicated via shared memory.
Early implementations have been optimized for specific alignment types and use cases~\cite{liu2013cudasw++,cudalign,korpar2013sw}.
More recent GPU approaches provide more flexible libraries and APIs, which are required for processing high-throughput \ac{NGS}
data such as GASAL\,2~\cite{gasal2}, ADEPT~\cite{ADEPT}, and
NVBIO~\cite{nvbio}. Wang et al.~\cite{wang2017communication} attempted to replace shared memory accesses in \ac{DP} computation by warp shuffles between registers but only achieved very moderate improvements due to an ineffective partitioning scheme. In addition, AnySeq~\cite{DBLP:conf/ipps/MullerS0MLKH20}
demonstrates that using PE and AnyDSL it is possible to build an alignment library based on higher-order abstractions that can specialize on a variety of architectures (CPUs, GPUs, and FPGAs).

This work, \anyseq, replaces the less efficient GPU kernels in AnySeq\,1.
It computes one alignment per warp thereby allowing for fast communication between thread-local registers by means of warp shuffles in systolic fashion.
This new parallelization with its associated partitioning scheme minimizes memory access overheads and achieves over 80\% of the theoretical peak performance of modern NVIDIA and AMD GPUs. It outperforms GASAL2, ADEPT, NVBIO, and the old AnySeq\,1 implementation by a factor of at least 3.6 and achieves a median speedup of 19.2$\times$ over these tools across different alignment scenarios and sequence lengths.

%==============================================================================
\section{Conclusion}\label{conclusion}
%==============================================================================

The continually increasing volume of genomic sequencing data has resulted in a growing demand for high-throughput sequence processing pipelines in recent years.
Sequence alignment is an important algorithmic part of many bioinformatics applications and a major contributor to their overall runtime.
% While the task of aligning batches of short reads still dominates many NGS pipelines, long read alignment with a high variance in individual read length is gaining in importance in recent years.

We have presented \anyseq, a high performance sequence alignment library that makes use of a new approach for implementing high-throughput DP algorithms on modern GPU architectures based on warp shuffles and half-precision arithmetic.
\anyseq's implementation achieves over 80\% peak performance on modern GPUs and outperforms state-of-the-art GPU-based competitors such as GASAL\,2, NVBIO and ADEPT by at least a factor of 6.6 and achieves a median speedup of 30$\times$ compared to these tools when running on the same hardware. \anyseq is also at least 3.6 times faster than the old AnySeq\,1 GPU implementation and achieves a median speedup of 7.6$\times$ compared to it (taking into account different alignment schemes, traceback and all tested read lengths).
Furthermore, \anyseq offers a clear advantage over state-of-the-art AVX512-enabled CPU libraries.
Using an AMD MI100 it outperforms SeqAn and AnySeq\,1 running on a dual-CPU workstation with 32 physical threads
by at least 9.8$\times$ with a median speedup of 13.2$\times$.

Our results show that prior GPU approaches have not been able to unlock the full potential of GPUs for alignment and only achieve a fraction of the available peak performance. As a consequence, they cannot provide significant speedups compared to current CPU-based approaches since they are bottlenecked by inefficient memory access schemes. Our approach thus changes the standing of GPUs in sequence alignment significantly. We demonstrate a GPU implementation that has close-to-peak performance on modern GPUs and a (median) 30-fold performance increase in terms of GCUPS compared to the best performing existing GPU codes for a variety of sequence lengths (ranging from 125 to 4,096 for Illumina reads, and 4,442 to 57,571 for PacBio reads). Our partitioning scheme allows for even longer sequence lengths while requiring only linear memory consumption. Thus, GPUs can be efficiently used for typical problem sizes used in practice for processing large-scale NGS datasets in bioinformatics.

Our parallelization scheme is in general not limited to pairwise alignment but is applicable to a wider range of DP-based algorithms with similar dependency relationships such as the Viterbi algorithm for finding a most likely sequence of hidden states in a hidden Markov model. It would thus be interesting to evaluate the performance of our approach when adapted to different \ac{DP} algorithms. Related speed advantages might become even more pronounced on upcoming GPUs such as the recently announced NVIDIA Hopper architecture by taking advantage of new DPX instructions for accelerating \ac{DP} algorithms \cite{hopper}.

Our implementation makes use of the AnyDSL compiler framework and is written in its functional front-end language Impala.
The partial evaluation guarantees provided by AnyDSL enable convenient higher-order abstractions for separating computation into common generic parts and parts that are optimized for specific hardware architectures.
Program parametrization for different alignment variants, scoring schemes and mappings to different architectures are achieved through simple function composition instead of complicated and hard-to-debug metaprogramming.
Performance tuning was greatly enhanced by the ability to trigger guaranteed specializations like loop unrolling instead of relying on unpredictable compiler heuristics. For the development of \anyseq, AnyDSL was particularly helpful to implement abstractions that made the creation and exploration of several algorithmic variants and parameter sets very easy. In a conventional setting, all these variants would have to be implemented manually. Using AnyDSL, they are generated automatically using PE without any runtime overhead.

\anyseq is open source software and can be downloaded at
\url{https://github.com/AnyDSL/anyseq}.

\begin{acks}
    This work is supported by the Federal Ministry of Education and Research (BMBF) as part of the HorME and MetaDL projects. %We acknowledge hardware access through NHR.
\end{acks}

%==============================================================================
\clearpage
\balance
\bibliographystyle{ACM-Reference-Format}
\bibliography{dblp,other}

\end{document}